\documentstyle[epsfig]{mn}
\begin{document}

\def\ltsima{$\; \buildrel < \over \sim \;$}
\def\simlt{\lower.5ex\hbox{\ltsima}}
\def\gtsima{$\; \buildrel > \over \sim \;$}
\def\simgt{\lower.5ex\hbox{\gtsima}}
\def\ls{{_<\atop^{\sim}}}
\def\lax{{_<\atop^{\sim}}}
\def\gs{{_>\atop^{\sim}}}
\def\gax{{_>\atop^{\sim}}}

\def\sax{{\it Beppo}SAX~}
\def \etal {et~al. }

\title[The 0.1-100 keV spectral shape and variability of Mkn421 in  
high state]
{The 0.1-100 keV spectrum and variability of Mkn421 in high state}

\author[A. Malizia et al.]{A. Malizia $^{1,2,3}$, M. Capalbi $^{1,3}$, 
F. Fiore $^{1,4}$, P. Giommi$^{1,5}$, G. Gandolfi $^{3,6}$, \\ ~ \\
{\LARGE A. Tesseri $^{3,6}$,
L. A. Antonelli $^{1,4}$, R. C. Butler$^{5}$, 
G. Celidonio$^{3,6}$, A. Coletta$^{3,6}$,  }\\ ~ \\  
{\LARGE L. Di Ciolo$^{3,6}$,
J. M. Muller$^{6,7}$, L. Piro$^{8}$, S. Rebecchi$^{3,6}$,   
D. Ricci$^{3,6}$, R. Ricci$^{3,6}$, } \\ ~ \\ 
{\LARGE M. Smith$^{6,7}$, 
V. Torroni$^{3,6}$ } \\ ~ \\ 
$^1$ \sax Science Data Center, Via Corcolle 19, I--00131 Roma, Italy \\
$^2$ Southampton University, SO17  1BJ, England \\
$^3$ Telespazio, Via Corcolle 19, I--00131 Roma, Italy \\
$^4$ Osservatorio Astronomico di Roma, Via dell'Osservatorio,
I--00044 Monteporzio Catone, Italy \\
$^5$ ASI, Area Ricerca Scientifica, Roma, Italy \\
$^6$ \sax SOC, Via Corcolle 19, I--00131 Roma, Italy \\
$^7$ SRON, Sorbonnelaan 2, 3584 CA Ultrecht, The Netherlands \\
$^8$ Istituto di Astrofisica Spaziale, CNR, Via Fosso Del Cavalire, I-00133,
     Roma, Italy\\
}
\maketitle
\begin{abstract}

The results of a \sax TOO observation of the BL Lac object Mkn421
during a high intensity state are reported and compared with
monitoring X-ray data collected with the \sax Wide Field Cameras (WFC)
and the RXTE All Sky Monitor(ASM).  The 0.1-100 keV spectrum of Mkn421
shows continuous convex curvature that can be interpreted as the
high-energy end of the synchrotron emission. The source shows
significant short-term temporal and spectral variability, which can be
interpreted in terms of synchrotron cooling. The comparison of our
results with those of previous observations when the source was a
factor 3-5 fainter shows evidence for strong spectral variability,
with the maximum of the synchrotron power shifting to higher energy
during high states. This behaviour suggest an increase in the number
of energetic electrons during high states.

\end{abstract}


\section{Introduction}

Mkn421 is a bright, nearby ($z$=0.03) BL Lac object classified as HBL
(High energy peaked BL Lac, Padovani \& Giommi 1996) since its
Spectral Energy Distribution (SED) peaks (in a $\nu f(\nu)~vs~\nu$
representation) in the UV/X-rays.  Mkn421 is one of a few
extragalactic objects (with Mkn501, 1ES2344+514 and PKS2155-304, all
HBLs) so far detected at TeV energies e.g. Punch et al. 1992 where
it shows tremendous variability, down to timescales of about 15
minutes (Gaidos et al. 1996). Mkn421 was repeatedly monitored with the
Whipple, HEGRA and CAT Cherenkov telescopes and, whenever possible,
simultaneously observed with X-ray satellites.  The X-ray emission
is also highly variable, with distinct differences between the soft
and hard X-rays.  Multi-wavelength campaigns have shown correlated
flux changes between the X-ray and the TeV region (Takahashi \etal
1996).

In this paper we report the results of a \sax (Boella et al. 1997a)
TOO observation of Mkn421 during a high intensity state.  The source
was hard and showed significant temporal and spectral variability on
timescales down to 500-1000 seconds.  A comparison with previous \sax
observations carried out when Mkn421 was less luminous also shows
remarkable spectral variations.  As seen in two other HBL BL Lacs,
namely Mkn501 (Pian et al. 1998) and 1ES2344+514 (Giommi et al. 1999),
both observed by \sax during high states, also in Mkn 421 the peak of
the synchrotron power significantly moves to higher energy, well into
the X-ray band.  These findings exploit the unique \sax spectral
coverage (from 0.1 up to 300 keV) to observe emission from the high
energy tail of the emitting electron distributions.

\section{Observation and Data analysis}

The \sax Narrow Field Instruments (NFI) observed Mkn421 on 22 June
1998 as part of a Target Of Opportunity (TOO) program dedicated to the
study of different types of AGN in high states.  Mkn421 was also
observed earlier by \sax on April and May 1997 as part of the normal
observation program.  The TOO observation was triggered when Mkn421
was detected in one of the \sax Wide Field Cameras (WFC, Jager et
al. 1997) at a flux level of $\sim$20 mCrabs, that is in a high state
compared with previous observations.  The WFC trigger is very
effective (as demonstrated by the \sax Gamma Ray Burst experience,
e.g. Costa 1998) since it allows a much faster response compared to
other triggering methods.

Figure 1 shows the X-ray light curve of Mkn421 during June 1998 (day
880 = May 31 1998). In the top panel filled circles represent one day
averages WFC measurements in the 2-10 keV energy band (4-7 orbits,
depending on the primary pointing). The error as given by the rms
between different orbits is 20\%.  The first point in the plot (open
circle) indicates the flux measured by the WFC during a single
available orbit and the uncertainty here is correspondingly higher.
The star indicates the flux seen during our observation of June 22,
1998.  The dashed lines identify the flux levels recorded during two
\sax NFI exposures on this source in 1997 (Guainazzi et al. 1998,
Fossati et al. 1998). The bottom panel shows the RXTE-ASM one day
average light curve in the 2-10 keV energy range (RXTE-ASM public
archive).  Variations of the order of a factor of two are evident on
time scales of a few days.  The peak near day 884 (= June 4 1998) in
the ASM light curve coincides with the largest flux recorded by the
\sax WFC.  The flux of about 40-50 mCrab in June 1998 is by far the
largest X-ray flux ever observed from Mkn 421.  Note that in June
1998, Mkn421 was always detected in a rather high state.

The \sax NFI used for the observation of Mkn 421 consist of a Low
Energy Concentrator Spectrometer (LECS, Parmar et al. 1997) sensitive
between 0.1 and 10 keV; three identical Medium Energy Concentrator
Spectrometers (MECS, Boella et al. 1997b) covering the 1.5-10 keV
band; and two co-aligned high energy instruments, the High Pressure
Scintillator Proportional Counter (HPGSPC, Manzo et al. 1997) and the
Phoswich Detector System (PDS, Frontera et al. 1997) operating in the
4-120 keV and 15-300 keV bands respectively.

The MECS was composed at launch by three identical units. On 1997 May
6$^{\it th}$ a technical failure caused the switch off of unit
MECS1. All observations after this date are performed with two units
(MECS2 and MECS3).  The LECS is operated during spacecraft dark time
only, therefore LECS exposure times are usually smaller than the MECS
ones by a factor 1.5-3.  The PDS and the HPGSPC are collimated
instruments with a FWHM of about 1.4 degrees. The PDS consists of four
phoswich units, and is normally operated with the two collimators in
rocking mode, that is with two phoswich units pointing at the source
while the other two monitor the background. The two halves are swapped
every 96 seconds. This default configuration was also used during our
observation of Mkn421.  The net source spectra have been obtained by
subtracting the off and the on counts.

The effective exposure time was 32516 seconds in the MECS, 11082
seconds in the LECS, 13912 in the PDS and 13750 in the HPGSPC.

Standard data reduction was performed using the software package
"SAXDAS" (see http://www.sdc.asi.it/software and the Cookbook for
BeppoSAX NFI spectral analysis, Fiore, Guainazzi \& Grandi 1998).
Data are linearized and cleaned from Earth occultation periods and
unwanted period of high particle background (satellite passages
through the South Atlantic Anomaly). The LECS, MECS and PDS
background is relatively low and stable (variations of at most 30 \%
during the orbit) thanks to the satellite low inclination orbit (3.95
degrees).  Data have been accumulated for Earth elevation angles $>5$
degrees and magnetic cut-off rigidity $>6$). For the PDS data we
adopted a fine energy and temperature dependent Rise Time selection,
which decreases the PDS background by $\sim 40 \%$. This improves the
signal to noise ratio of faint sources by about 1.5 (Frontera et
al. 1997, Perola et al. 1997, Fiore, Guainazzi \& Grandi 1998).  Data
from the four PDS units and the two MECS units have been merged after
equalization and single MECS and PDS spectra have been accumulated.
We extracted spectra from the LECS and MECS using 8 arcmin radius
regions.  LECS and MECS background spectra were extracted from blank
sky fields from regions of the same size in detector coordinates.

\begin{table*}
\centering
\caption{Mkn421 spectral fits} 
\begin{tabular}{lcccccc}
\hline \hline
model (0.1-100 keV)   &  $\alpha_E$ & $\alpha_H$ & $E_{0}^a$ & $\beta^b$
& $E_{f}^{a,c}$ & $\chi^2$ (d.o.f.)\\
\hline
April 1997 & &&\\
PL+ABS$^{\star}$           & 2.4  & --  & --   &      & -- & 2086 (136) \\
CurvedPL+ABS$^{\star}$   & 1.4  & 2.7 & 3.91 & 0.32 & -- & 112.8 (136) \\
\hline
May 1997 & &&  \\
PL+ABS$^{\star}$           & 2.4  & --  & --   & -- & -- & 921.8 (136) \\
CurvedPL+ABS$^{\star}$   & 2.0    & 2.8 & 3.89 & 0.84 & -- & 155.0 (136) \\ 
\hline
June 1998 & &&\\
PL+ABS$^{\star}$           & 2.2  & --  & --  & --   & --   & 4027 (201) \\
CurvedPL+ABS$^{\star}$     & 1.8  & 2.4 & 2.7 & 1.08 & --  & 291 (201) \\
CurvedPL+ABS$^{\star}$+Highecut  & 1.8 & 2.3 & 1.72  & 1.1 & 30 & 239 
(201) \\ 
\hline \hline

\end{tabular}

$^{\star}N_H$ = 1.6 $\times$ $10^{20}$ cm$^{-2}$ FIXED; $^a$ in keV; 
$^b$ = curvature radius; $^c$ Folding Energy
\end{table*}

\begin{figure}
\epsfig{ file = 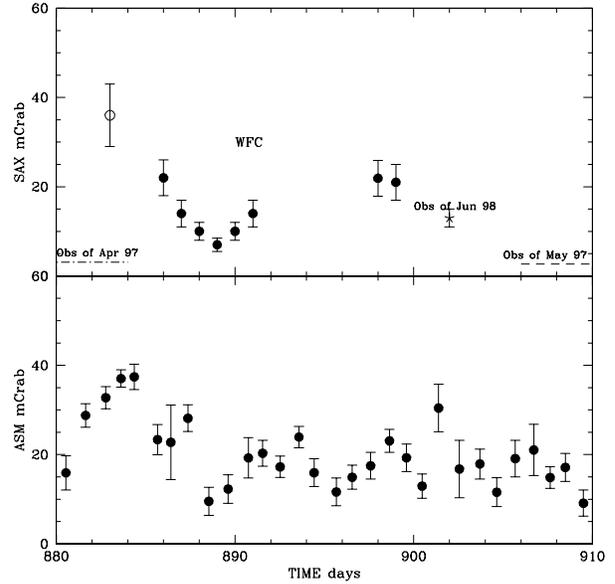, height=8.5cm}
\caption{\sax-WFC light curve (top panel) compared with 
ROSSI-ASM one day average (bottom panel) during \sax observations of 
June 1998.} 
\end{figure}

\section{Results}

\subsection{Temporal Analysis}

\begin{figure}
\epsfig{ file =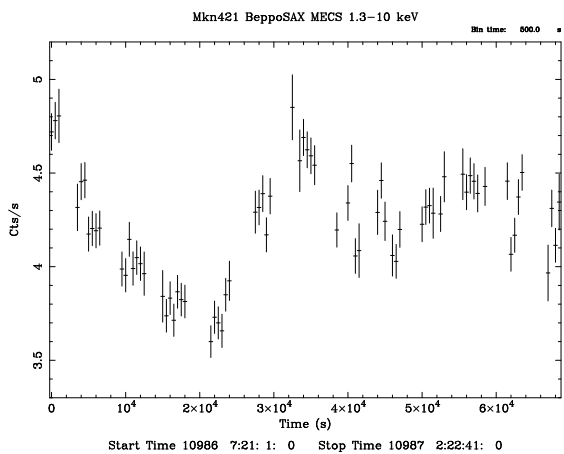, height=8.5cm, angle=-90 }
\caption{MECS light curve (1.3--10) keV with a binning time of 500s}
\end{figure}

Figure 2 shows the MECS 1.3--10 keV light curve in bins of 500
seconds.  Variations of a factor of 25\% are present on time-scales of
8-10 ks (see the rise from time=24ks to time=32ks). Variations of a
smaller amplitude are present down to a 500-1000 seconds timescale
(see for example the events at time=24 ks, 46ks, 62ks in Figure 2).
These variations are accompanied by strong spectral variability.  This
is shown in Figure 3, where we plot the light curves in the PDS 13-50
keV band, LECS 0.1-0.7 band, MECS ``Soft'' 1.3-3 keV and ``hard'' 5-10
keV bands, along with the MECS Hard/Soft hardness ratio (HR) (from top
to bottom).  The bin size used, 2850 seconds, roughly corresponds to
half of the satellite orbit, and helps in illustrating how the LECS
data are actually acquired for part of each orbit only. Therefore,
they are not completely simultaneous to the MECS and PDS data. This
can introduce an offset between the normalization in the LECS and the
other instruments in spectral fits (see next section).  The comparison
of the light curves in different energy bands (in particular the two
MECS light curves and the LECS light curve) shows that the source
hardens while brightening, in agreement with the results of Giommi et
al. (1990), Takahashi et al (1996) and Sambruna et al. (1996).
Unfortunately, the statistics in the PDS is not good enough to allow
us to search for an extension of this trend at energies higher than 10
keV.  The behaviour of the MECS HR suggests that the hard X-rays lead
the soft X-rays.  Figure 4 shows the 5-10 keV/1.3-3 keV hardness ratio
HR plotted against the intensity (1.3-3keV + 5-10keV count rate), in
bins of 5700 seconds (roughly one orbit).  Numbers indicate
progressive orbits. Starting from orbit number 1, the HR first
decreases with decreasing count rate, and successively increases
again, but at an higher rate, following a clockwise motion.  Takahashi
et al. (1996) found a similar behaviour in an ASCA observation, when
the source was at a flux level similar or slightly higher than during
our \sax TOO observation, indicating again that hard X-rays lead the
soft X-rays.  To measure the lag time we have calculated the Discrete
Correlation Function (Edelson \& Krolik 1988) between the MECS Soft
and Hard light curves. We find a lag of 1700$\pm$600 seconds (90 \%
confidence interval).

\begin{figure}
\epsfig{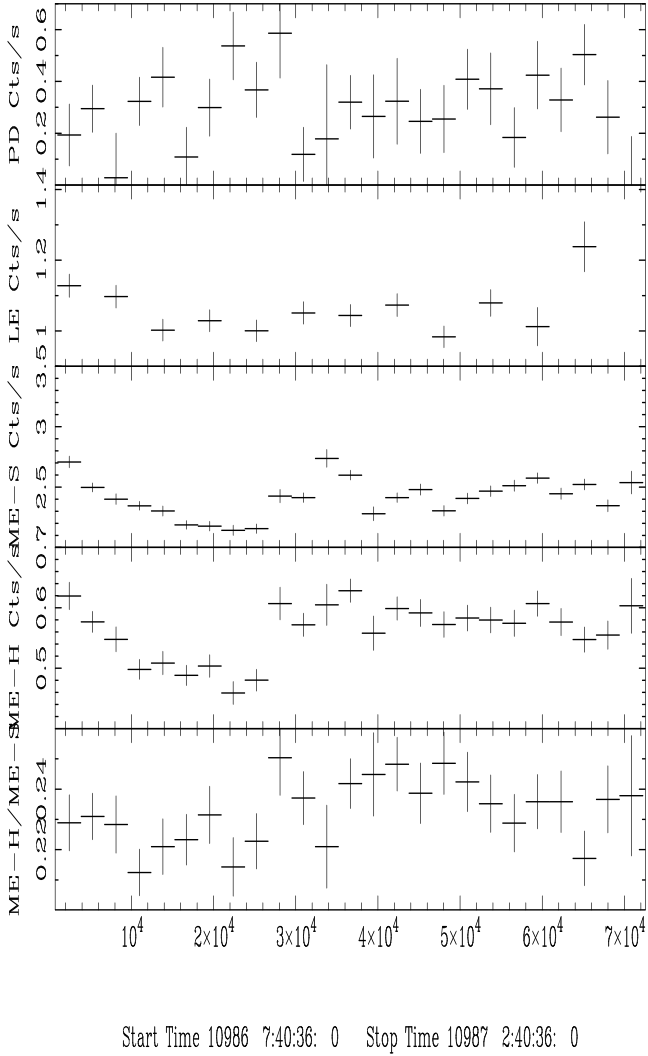}
\caption{Light curves in the 13--50 keV (PD), 0.1--0.7 keV (LE), 1.3--3
keV (ME-H) and 5--10 keV (ME-S) and their hardness ratio (ME-H/ME-S)}
\end{figure}

\begin{figure}
\epsfig{ file =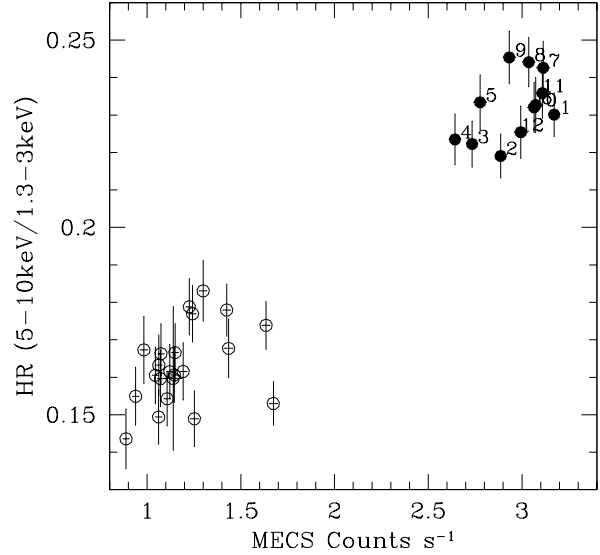, height=8.5cm}
\caption{Hardness ratio (5-10 keV/1.3-3 keV) 
versus intensity (1.3-3 keV + 5-10 keV). 
In the left-bottom corner the April 97 counts have been plotted  
while the TOO counts are the points located in the right-top corner.
The numbers put on the side of our data points indicate progressive
orbit (5700s)}
\end{figure}

\subsection{Spectral Analysis}

Spectral fits were performed using the XSPEC 9.0 software package and
public response matrices as from the 1998 November issue.  PI channels
are rebinned sampling the instrument resolution with the same number
of channels at all energies when possible and to have at least 20
counts per bin.  This guarantees the use of the $\chi^2$ method in
determining the best fit parameters, since the distribution in each
channel can be considered Gaussian.

Constant factors have been introduced in the fitting models in order
to take into account the intercalibration systematic uncertainties
between instruments (Cusumano et al. 1999, Fiore, Guainazzi \& Grandi
1998). The expected factor between LECS and MECS is about 0.9. In the
fits we use the MECS as reference instruments and constrained the LECS
parameter to vary in the small range 0.8-1.  The expected factor for
MECS and PDS is 0.8 and we constrained the PDS parameter to vary in
the range 0.7-0.9.

The energy range used for the fits are: 0.1--4 keV for the LECS
(channels 11--400), 1.65--10 keV for the MECS (channels 37--220) and
13--100 keV for the PDS.

Many existing narrow band X-ray spectra of blazars are sufficiently
well described by a single power-law model. However recent wide band
X-ray spectra of Mkn421 (Fossati et al. 1998, Guainazzi et al. 1998),
PKS 2155-304 and other bright BL Lacs require more complex models like
a broken power-law or a curved spectrum (e.g. Giommi et al. 1998,
Wolter et al. 1998).

Figure 5 shows the \sax LECS, MECS, HPGSPC, PDS, 0.1-100 keV spectrum
of Mkn421 measured during the June 1998 observation and fitted to a
simple power-law plus low energy absorption due to a $N_H$ column
equal to the Galactic value along the line of sight.  A single
power-law model is clearly an unacceptable representation of the data
($\chi{^2}_{\nu}$ = 23).  Figure 5 demonstrates that this is not due
to a localized feature but to an incorrect modelling of the spectrum
across the entire 0.1-100 keV energy range. In fact, the residuals plotted
at the bottom of figure show that large convex spectral curvature is
present.

A gradual steepening with energy is in line with the Synchrotron Self
Compton (SSC) mechanism, a widely accepted scenario to explain the SED of
HBL objects (Ghisellini et al. 1998).  We have thus fitted our data to
a {\it curved spectrum} (Matt, private communication) defined as
follows

$$ F(E)=E^{-[(1-f(E)*\alpha_E + f(E)*\alpha_H]} $$

where f(E)=[1-exp(-E/E$_{0}$)]$^{\beta}$, $\alpha_E$ and $\alpha_H$
are the low and high energy asymptotic energy indices E$_{0}$ is a
break energy and $\beta$ is the curvature radius.  The column density
has been fixed to the Galactic value along the line of sight of Mkn421
($N_H$ = $1.6 \times 10^{20} cm^{-2}$, Dickey \& Lockman 1990).  This
model has been successfully applied to the previous \sax observation
of Mkn421 (Guainazzi et al. 1998) and PKS2155-304 (Giommi et
al. 1998).

This curved model gives acceptable fits to the 1997 \sax
observations. However the June 1998 spectrum requires additional
curvature; a good fit ($\chi^{2}$=189/153) can be obtained adding a
high energy cutoff to the model (see figure 6).  The analysis of the
residuals in figure 6 shows a deviation of the order of 30\% below 0.5
keV, probably due to the carbon edge like feature in the LECS energy
range. Moreover at lowest energies (0.1-0.2 keV) a more curved model
seems to be required.  The results of our spectral analysis on all the
observations considered in this paper are summarized in Table1.  Our
results on the April 1997 observation of Mkn421 (taken from the \sax
public archive) are well in agreement with the original analysis
presented in Guainazzi et al. (1998).

\begin{figure}
\epsfig{ file = 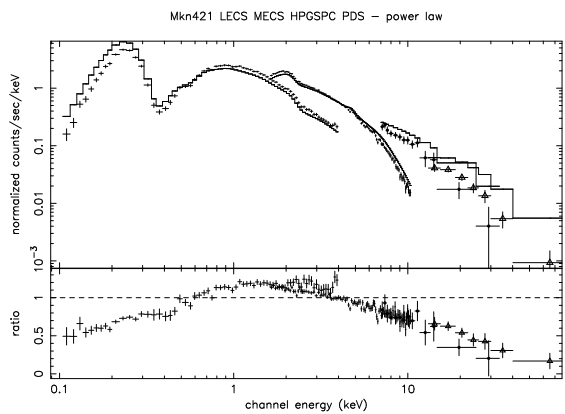, height=8.5cm, angle=-90}
\caption{\sax broad band spectrum of Mkn421 during 
the TOO of June fitted with a simple power law. The filled circles
and the triangles represent the HPGSPC and PDS data respectively.} 
\end{figure}

\begin{figure}
\epsfig{ file = 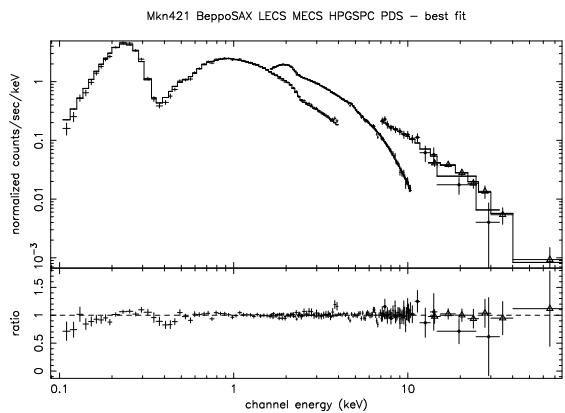, height=8.5cm, angle=-90}
\caption{\sax broad band spectrum of Mkn421 during 
the TOO of June fitted with a curved power law plus a high energy cutoff.
The filled circles and the triangles represent the HPGSPC and PDS data 
respectively.} 
\end{figure}

\section{Comparison with previous observations}

\sax has observed Mkn421 in several campaigns between 1997 and 1998
and in particular the April 1997 and May 1997 observations have been
taken into consideration in the present work to study the spectral
variations of the source.

Figure 4 compares the MECS HR measured during the June 1998
observation with that measured during three April 1997 observations,
when the source was in a quiescent state (Guainazzi et al. 1998).
During the April 1997 observations the HR increases with increasing
intensity until it saturates (Guainazzi et al 1998), in line with what
was found in a similar quiescent state by Giommi et al. 1990 and
Sambruna et al. 1994.  The HR in June 1998 is much higher than during
the 1997 observations, indicating that the saturation possibly
concerns single variability events only. The 1998 HR does not
saturate, indicating that either the observation did not catch the
source at the peak of a variability cycle, or that HR saturation does
not apply to variability events in high source states.

The comparison of the June 1998-TOO observation of Mkn421 with the
1997 \sax observations shows strong spectral variations.  In figure 7
we have plotted the ratio between the spectrum seen during our TOO
observation and that during the April 97 (open squares) and May 97
(filled circles) observations.  It is evident that the source hardened
significantly when it brightened (up to a spectral ratio of 4-5 at 10
keV).  This hardening is more pronounced above 1 keV. Below this
energy the spectral ratio is less than a factor 2 showing that most of
the variability occurred at high energy.

In figure 8 we report the 0.1-100 keV spectra of Mkn 421 during the
three observations considered, multiplied by the frequency $\nu $. The
maxima in this plot identify the region where most of the synchrotron
power is emitted.  During the high state the peak is located at
log$\nu \sim 17.4$, or about 1 keV, while during the other
observations the peak was below 1 keV.

\section{Discussion and Conclusions}

The spectral energy distribution of high-energy peaked BL Lacs, from
radio to $\gamma$-ray energy, in the $\nu$-$\nu$F$_{\nu}$
representation, is characterized by two peaks: one in the UV/soft
X-ray band and the second at the GeV to TeV energies.  This spectral
energy distribution is generally interpreted as due to incoherent
synchrotron radiation followed by Inverse Compton emission
(e.g. Ghisellini, Maraschi \& Dondi 1996).  The radio to X-ray
emission is produced by the synchrotron process as strongly suggested
by the connection of the X-ray and IR, optical, UV spectra.  On the
contrary the inverse Compton is responsible for the $\gamma$-ray
emission and the correlated flaring at X-ray and TeV energies
(Takahashi et al. 1996).  Many theoretical models have been proposed
to explain the spectral and the timing variability observed in the
high-energy peaked BL Lac.  The comparison between different spectral
states can give us information on the source electron distribution.
The 1997 and the 1998 \sax data show both an increase in power and a
shift in the synchroton peak. This suggests that during high states
the number of energetic electrons is higher and that a simple increase
of the magnetic field or of the electron cutoff energy is not
sufficient to explain the data.  A possible explanation could be the
injection of energetic electrons caused by shocks in a relativistic
jet (Kirk et al. 1998).  A behavior similar to that of Mkn421 has been
observed in Mkn 501 (Pian et al. 1998, Ghisellini 1998) and
1ES2344+514 (Giommi et al. 1999).

On the other hand, the study of the short term variability can provide
detailed information on the cooling mechanisms and the source
geometry.  The characteristic short time scale change of the hardness
ratio as a function of the count rate and the lags between the hard
and soft photons have been interpreted by Takahashi et al. (1996) in
terms of Syncrotron cooling.  They estimated a time lag between hard
(2-7.5 keV) and soft (0.5-1 keV) energy bands of the order of one hour
during an ASCA observation when the source was in a state similar to
that seen during our \sax TOO observation.  Following Takahashi et
al. (1996) and assuming a t$_{sync}$ $\sim$ 1.2 $\times$ 10$^{3}$
B$^{-3/2}$E$_{keV}^{-1/2}$ $\delta^{-1/2}$, where E$_{keV}$ is the
observed energy, using their values of B $\sim$ 0.2 G and $\delta$ = 5
we have calculated that the time lag between hard (5-10 keV) and soft
(1.3-3.5 keV) emission in our observation should be of about 1500
seconds ($\sim$ 25 min) which is consistent with the value found with
the DCF method (see section 3.1).

In a recent work Chiaberge and Ghisellini (1999) studied the time
dependent behaviour of the electron distribution injected in the
emitting region and proposed a comprehensive model to describe the
evolution of synchrotron and self Compton spectra. They pointed out
that the cooling time can be shorter that the light crossing time
(R/c) and if this is the case the particle distribution will evolve
more rapidly than R/c and the observer will see the contribution of
the different spectra produced in each {\it slice} of the source.
Taking into account the light time crossing effects, the different
cooling times of electrons emitting at various frequencies can cause
remarkable time delays, of the order of those observed by us and by
Takahashi et al. (1996).

\begin{figure}
\epsfig{file=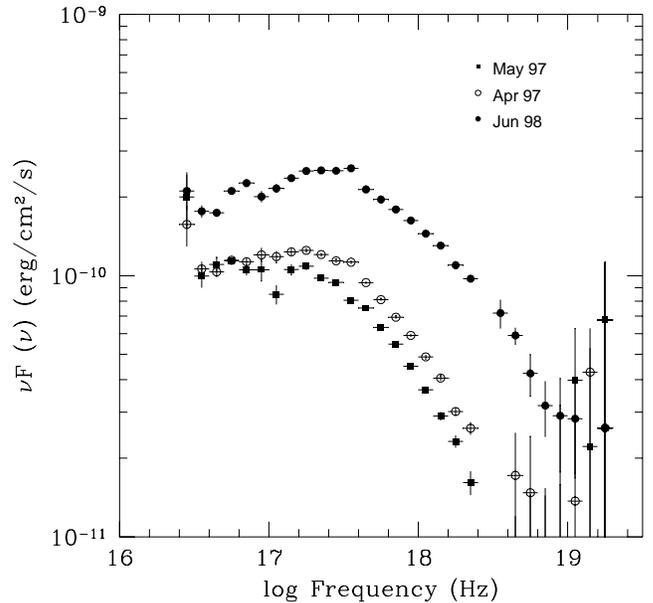, height=9.5cm}
\caption{Spectral ratio of the June 1998 TOO observation to those of 
April 97 (open squares) and May 97 
(filled circles). 
Both ratios increase with energy, reaching a maximum of
4-5 at 10 keV.}
\end{figure}

\begin{figure}
\epsfig{file=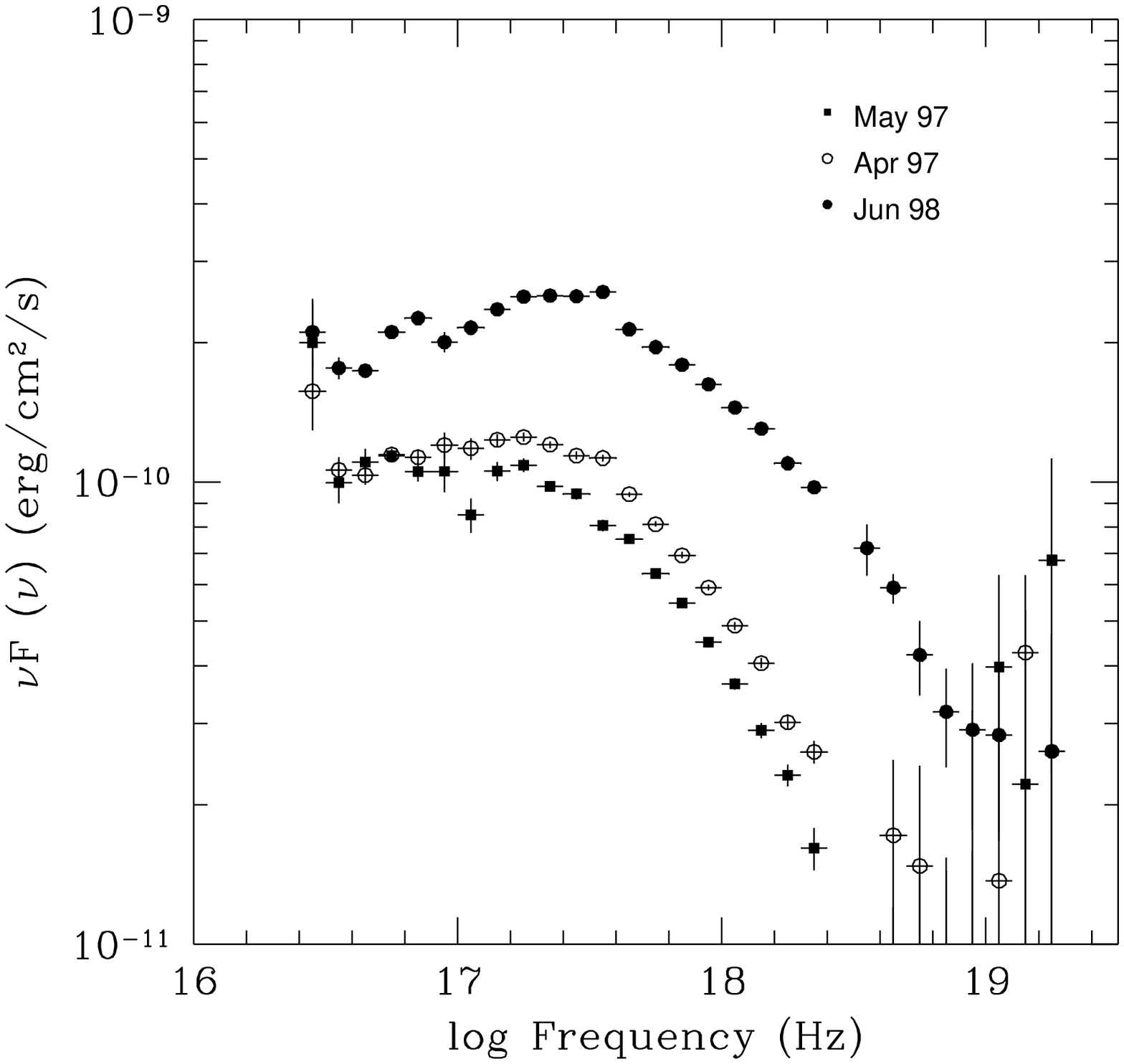, height=9.5cm}
\caption{0.01-100 keV spectra of Mkn421 during the three
\sax observations, multiplied by the frequency.}
\end{figure}

\bigskip
We thank G. Matt for providing the XSPEC gradually 
changing index power-law model used to fit the data in this paper.

\end{document}